\documentclass[preprints,article,accept,moreauthors,pdftex]{Definitions/mdpi}

\firstpage{1} 
\pubvolume{1}
\issuenum{1}
\articlenumber{0}
\pubyear{2022}
\copyrightyear{2022}
\datereceived{} 
\dateaccepted{} 
\datepublished{} 
\hreflink{https://doi.org/} 
\pdfoutput=1

\Title{QUADRATIC THEORY OF GRAVITY WITH A SCALAR FIELD AND TYPE I SHAPOVALOV WAVE SPACETIMES}

\TitleCitation{Quadratic theory of gravity with a scalar field and type I Shapovalov wave spacetimes}

\Author{
Konstantin Osetrin $^{1, 2,
\ddagger}$\orcidA{0000-0003-4739-2788}, 
Ilya Kirnos $^{1,\ddagger}$ \orcidB{0000-0003-3737-3583}, 
and 
Altair Filippov $^{1,\ddagger}$ \orcidC{0000-0001-5998-2238}
}

\AuthorNames{Konstantin Osetrin, Ilya Kirnos and Altair Filippov}

\AuthorCitation{Osetrin, K.E.; Kirnos, I.V.; Filippov, A.E.}

\address{%
$^{1}$ \quad Tomsk State Pedagogical University; osetrin@tspu.edu.ru\\
$^{2}$ \quad Tomsk State University; osetrin@gmail.com}

\secondnote{These authors contributed equally to this work.}

\abstract{
For the quadratic theory of gravity with a scalar field, exact solutions are found for gravitational-wave models in Shapovalov~I~type spacetimes, which do not arise in models of the general theory of relativity.
The theory of gravity under consideration can effectively describe the early stages of the dynamics of the universe. Type I Shapovalov spaces are the most general form of gravitational-wave Shapovalov spacetimes, whose metric in privileged coordinate systems depends on three variables, including the wave variable. For Einstein vacuum spacetimes, these wave models degenerate into simpler types. The obtained exact models of gravitational waves in the quadratic theory of gravity can be used to test the realism of such theories of gravity.
}

\keyword{quadratic theory of gravity; gravitational waves; Hamilton-Jacobi equation; eikonal equation; Killing fields; Shapovalov spacetimes; test particle trajectories.} 

\MSC{83C10, 83C35}
\begin{document}
%

\section{Introduction}

In this paper, we consider the problem of constructing models of gravitational waves as exact solutions of field equations in the framework of the quadratic theory of gravity with a scalar field based on Shapovalov type I wave spacetimes.

The first surge of interest in scalar-tensor theories of gravity is apparently associated with the emergence of the Brans-Dicke theory \cite{Brans-Dicke1961}. Guided by the Mach principle, they proposed the Lagrangian of the gravitational field of the form
\begin{equation}
\label{lagrBD1}
\mathcal{L}=
\varphi R
-\omega\displaystyle\frac{\partial_\mu\varphi\partial^\mu\varphi}{\varphi}
+\displaystyle\frac{16\pi}{c^4}\mathcal{L}_{Matter}
,\end{equation}
where $R$ is the scalar curvature, $\varphi$ is the scalar field, $\omega$ is a constant.

Subsequently, other researchers generalized this Lagrangian to the form (see, for example, \cite{Carroll2004}):
\begin{equation}
\label{lagrBD2}\mathcal{L}=
f(\varphi)R
-\displaystyle\frac{1}{2}h(\varphi)\,\partial_\mu\varphi\partial^\mu\varphi-U(\varphi)
+\displaystyle\frac{16\pi}{c^4}\mathcal{L}_{Matter}
,\end{equation}
where $f(\varphi)$, $h(\varphi)$, $U(\varphi)$ are arbitrary functions of the scalar field $\varphi$.

As for theories quadratic in curvature, on the one hand, they can be obtained as a purely mathematical generalization of Einstein's gravity for the case of non-linearity of the gravitational field equations in second derivatives \cite{Lovelock1971}. On the other hand, quadratic theories arise as low-energy expansions in string theory, quantum gravity, M-theory \cite{Boulware1985}. They proved to be useful when considering the early Universe \cite{Mandal2021, Fomin2020}, especially the inflationary phase (Starobinskii's inflation \cite{Starobinskii1979} was based on $R^2$ gravity). However, at present, inflationary models with a scalar field - "inflaton" \cite{Weinberg2008, Gorbunov-Rubakov2010}, are more common, the first of which should be considered the Guth model \cite{Guth1981}. In this regard, it is of interest to consider the early Universe within the framework of combining these theories, where the Lagrangian contains both terms quadratic in curvature and a scalar field.

As an additional justification for this approach, it should be pointed out that purely Lovelock corrections to Einsteinian gravity (the theory with second-order corrections is known as the Einstein-Gauss-Bonnet theory) contribute to the equations of motion only in spaces of dimension above 4 (as an example of such a consideration see for example \cite{Ernazarov-Ivashchuk2022}). Inflation in quadratic theory with a scalar field was studied, for example, in \cite{Fomin-Chervon2019}

An overview of various modifications of general relativity and their comparison with cosmographic tests is given in \cite{Bamba2012}. Also, a large review of theories with Lagrangians that are nonlinear in curvature and contain the Gauss-Bonnet invariant is given in \cite{Nojiri2011}. Theories with Lagrangians of the form $f(\varphi,R)$ and their relation to inflation and modern accelerated expansion are considered in \cite{Nojiri2017}.

To describe the early stages of the Universe, spatially homogeneous non-isotropic models of spacetime are considered more realistic models than homogeneous and isotropic Friedman-Robertson-Walker models, especially since in many models \cite{Heckmann1989, Khalatnikov2003} non-isotropy decreases with time, space becomes isotropic \cite{Dynamical1997}.

Finally, the description of primordial gravitational perturbations (primordial  gravitational waves), the occurrence of which is predicted at the early (quantum) stages of the development of the Universe, requires the use of adequate mathematical methods to describe such ''wave'' models of spacetime.

It should be noted that not in all theories of this kind, the speed of a gravitational wave and the speed of light in vacuum coincide. Meanwhile, the gravitational wave burst GW170817, associated with the merger of neutron stars \cite{Abbott2017PRL161101}, was also accompanied by electromagnetic radiation, and the burst of electromagnetic radiation came somewhat later, which is associated with the physics of the process and, in addition, with the delay of electromagnetic radiation by the interstellar medium. Thus, it can be argued that in a vacuum the velocities of gravitational and electromagnetic waves coincide. The question of the resulting restrictions on the form of the Lagrangian of a theory quadratic in curvature is studied in \cite{Jana2021, Odintsov2021, Oikonomou2020}.

As adequate mathematical models for the physical problems under consideration, we propose to use Shapovalov wave-like spacetimes \cite{Osetrin2020Symmetry}. These spaces allow the existence of "privileged" coordinate systems, where it is possible to completely separate the variables in the equations of motion of test particles in the Hamilton-Jacobi formalism with the separation of wave variables, on which the spacetime metric depends in privileged coordinate systems (wave-like spacetimes).

The presence of such wave solutions makes it possible to form tests for observational checks of the realism of such models, since the presence of such types of gravitational waves could be reflected in the microwave background of the universe or in the stochastic gravitational wave noise observed during the detection of gravitational waves.

\section{
Field equations of quadratic gravity theory with scalar field
}

The Lagrangian of the theory of gravity we are considering has the following form:
\begin{equation}
\mathcal{L}=
\displaystyle
\frac{\sigma(\varphi)R+\gamma(\varphi)R^2}{2\varkappa^2}
-\xi(\varphi)\mathcal{G}
-\frac{1}{2}\omega g^{\mu\nu}\partial_\mu\varphi\partial_\nu\varphi
-V(\varphi)
+\mathcal{L}_{Matter},
\end{equation}
where $R$ is the scalar curvature, $\mathcal{G}$ is the Gauss-Bonnet term, $\varkappa$ is the gravitational constant, $\omega$ is the scalar field constant, $V(\varphi)$ is the scalar field potential, $\mathcal{L}_{Matter}$ is the Lagrangian of matter.

Then, by varying the Lagrangian, we obtain the field equations of this theory in the following form:
\begin{equation}
Q_{\mu\nu}=T_{\mu\nu}
,\qquad
\alpha,\beta,\gamma,\mu,\nu=0,...3
\label{FieldEquations1}
,\end{equation}
\begin{eqnarray}
Q_{\mu\nu}&\equiv&
\displaystyle\frac{1}{\varkappa^2}
\Bigl(
\left[\sigma(\varphi)+2\gamma(\varphi)R\right]R_{\mu\nu}
-\nabla_\mu\nabla_\nu\left[\sigma(\varphi)
+2\gamma(\varphi)R\right]
+g_{\mu\nu}\Box\left[\sigma(\varphi)
+2\gamma(\varphi)R\right]
\Bigr)
\nonumber\\&&
\vphantom{\frac{1}{2}}
-8\nabla_\alpha\nabla_\beta\left[\xi(\varphi) R_{\mu\phantom{\alpha}\nu}^{\phantom{\mu}\alpha\phantom{\nu}\beta}\right]
- 8\nabla_\gamma\nabla_\nu\left[\xi(\varphi){R^\gamma}_\mu\right]
-8\nabla_\gamma\nabla_\mu\left[\xi(\varphi){R^\gamma}_\nu\right]
\nonumber\\&&
\vphantom{\frac{1}{2}} 
+ 8\Box\left[\xi(\varphi)R_{\mu\nu}\right]
+ 8g_{\mu\nu}\nabla_\alpha\nabla_\beta\left[\xi(\varphi)R^{\alpha\beta}\right]
+ 4\nabla_\mu\nabla_\nu\left[\xi(\varphi)R\right]
-4g_{\mu\nu}\Box\left[\xi(\varphi)R\right]
\nonumber\\&&
\vphantom{\frac{1}{2}}
+16\xi(\varphi)R_{\alpha\mu}{R^\alpha}_\nu
- 4\xi(\varphi)RR_{\mu\nu}
- 4\xi(\varphi)R_{\mu\alpha\beta\gamma}{R_\nu}^{\alpha\beta\gamma}
\nonumber\\&&
\vphantom{\frac{1}{2}}
-\omega\partial_\mu\varphi\partial_\nu\varphi
+\frac{\omega}{2} g_{\mu\nu} g^{\alpha\beta}\partial_\alpha\varphi\partial_\beta\varphi
\nonumber\\&&
- \displaystyle\frac{\sigma(\varphi)R
+\gamma(\varphi)R^2}{2\varkappa^2}g_{\mu\nu}
+ \xi(\varphi)\mathcal{G}g_{\mu\nu}
+V(\varphi)g_{\mu\nu} 
\label{FieldEquations2}
,
\end{eqnarray}
where $T_{\mu\nu}$ is the energy-momentum tensor, $\nabla_\mu$ is the covariant derivative,
$\Box=g^{\alpha\beta}\nabla_\alpha\nabla_\beta$.

We obtain the equation for the scalar field in the following form:
\begin{equation}
\omega\,\Box\varphi+\displaystyle\frac{\sigma'(\varphi)R+\gamma'(\varphi)R^2}{2\varkappa^2}
- \xi'(\varphi)\,\mathcal{G}-V'(\varphi)=0,
\label{ScalarEquation}
\end{equation}
where the top prime means the ordinary derivative with respect to the variable on which the function depends.

We will consider the energy-momentum tensor of pure radiation:
\begin{equation}
T_{\mu\nu}=\varepsilon L_\mu L_\nu
,\qquad
g^{\mu\nu}L_\mu L_\nu=0
,
\end{equation}
where $\varepsilon$ is the radiation energy density, $ L_\mu$ is the radiation wave vector.

\section{
Shapovalov's wave models of spacetime
}

Shapovalov spaces \cite{Osetrin2020Symmetry} are gravitational-wave spacetime models that arise when constructing exact solutions for the equations of motion of test particles in the Hamilton-Jacobi formalism and for the eikonal equation (radiation propagation) by the method of separation of variables with the selection of wave variables along which the spacetime interval vanishes.
The speed of propagation of gravitational waves in observations of gravitational and electromagnetic radiation during the merger of neutron stars coincides with the speed of light \cite{Abbott2017PRL161101}, which experimentally confirms the use of wave variables along which the spacetime interval vanishes.
For the 4-dimensional case of spacetime, there are three main types of Shapovalov spacetimes according to the number of commuting Killing vectors they allow, which are included in the so-called comlete set of integrals of motion for the Hamilton-Jacobi equation. These spacetimes allow the existence of privileged coordinate systems, where separation of variables is possible in the Hamilton-Jacobi equation and in the eikonal equation. Type I Shapovalov wave spacetimes admit one Killing vector and, accordingly, in the privileged coordinate system their metric depends on three variables, including the wave variable. For vacuum Einstein equations, type I Shapovalov spaces lead to degeneration and to the appearance of additional commuting Killing vectors with a decrease in the number of ignored variables on which the metric depends. Nevertheless, in the presence of an additional electromagnetic field, solutions of the Einstein equations for Shapovalov spacetimes of type I arise, and the wave variable enters into the conformal factor of the metric. When considering modified theories of gravity, gravitational-wave solutions could also arise for type I Shapovalov wave spacetimes.

Let us consider gravitational-wave models of Shapovalov spaces of type I, for which the metric can be written in the privileged coordinate system in the following form:
\begin{equation}
g^{\alpha\beta}={f_0}
\left(
\begin{array}{cccc}
0&1&0&0\\
1&0&0&0\\
0&0&\frac{1}{W}&0\\
0&0&0&\frac{1}{W}
\end{array}
\right)
\label{metric1}
,\end{equation}
where
\begin{equation}
f_0=f_0(x^0),\ \ \ \ W(x^2,x^3)=t_3(x^3)-t_2(x^2)
.\end{equation}
In the privileged coordinate system used, the variables $x^0$ and $x^1$ are null variables along which the spacetime interval vanishes.

The considered models of spacetime for the Hamilton-Jacobi equation of test particles
\begin{equation}
g^{\alpha\beta}\frac{\partial S}{\partial x^\alpha}\frac{\partial S}{\partial x^\beta}=m^2c^2
,\qquad
\alpha,\beta,\gamma=0,1,2,3.
\label{HJE}
\end{equation}
allow finding the complete integral of this equation by the method of separation of variables, which in turn allows obtaining exact solutions for geodesic deviation equations and finding the explicit form of tidal accelerations in exact models of gravitational waves \cite{Osetrin2022EPJP856,Osetrin2022894}.
In the equation (\ref{HJE}) $S$ is the action function of the test particles, $m$ is the mass of the particle, $c$ is the speed of light, which we will choose as unity below.

Previously, we obtained exact models of gravitational waves with pure radiation related to Shapovalov spacetimes of type I in Einstein's theory \cite{OsetrinVaidya1996} and a number of wave solutions of other types in scalar-tensor and quadratic theories of gravity \cite{OsetrinScalar20181383,OsetrinScalar20202050275,OsetrinSymmetry2021,%
OsetrinScalar3120202050184,Osetrin2021092501}.

Note here that the scalar Klein-Gordon equation admits integration by the method of separation of variables in the same privileged coordinate systems as the Hamilton-Jacobi equation \cite{Shapovalov1978I, Shapovalov1978II}.

\section{
Analysis and solution of field equations
}

In this work, we consider the possibility of constructing exact models of gravitational waves in quadratic theories of gravity with a scalar field and radiation based on Shapovalov spacetimes of type I.
The general properties of the energy-momentum tensor of pure radiation in wave spacetime models that allow separation of variables in the eikonal equation were considered by us earlier in \cite{OsetrinRadiation2017}.
Wave models were considered earlier for Shapovalov spacetimes of type II \cite{OsetrinScalar20181383,OsetrinScalar20202050275,OsetrinSymmetry2021} and for Shapovalov spacetimes of type III \cite{OsetrinScalar3120202050184}, where exact models of gravitational waves in scalar-tensor theories of gravity were constructed.

The use in Shapovalov spacetimes of privileged coordinate systems with two null variables, along which the spacetime interval vanishes, suggests considering models where the scalar field can depend on both null variables.

Since the privileged coordinate system used allows the integration of the equations of the eikonal equation, the Hamilton-Jacobi equation for test particles, and the scalar Klein-Gordon equation by separation of variables,
we will look for exact solutions of the field equations for the Shapovalov wave spacetimes, assuming that the scalar field depends on the null variables $x^0$ and $x^1$ in a separated form:
\begin{equation}
\phi=\phi_0(x^0)\,\phi_1(x^1).
\end{equation}
The wave vector of radiation in the privileged coordinate system used, due to field equations and the normalization condition, takes the form:
\begin{equation}
L_{\mu}=\Bigl\{L_0,0,0,0 \Bigr\}
.
\end{equation}

The field equation $Q_{23}=0$ from the system of equations (\ref{FieldEquations1}) gives the condition:
$$
{\gamma}(\phi)
\biggl(
{t_2}''' {t_3}' ({t_2}{}-{t_3}{})^2
+{t_3}''' {t_2}'{} ({t_2}{}-{t_3}{})^2
$$
\begin{equation}
\mbox{}
+6 {t_2}'{}{t_3}'
\Bigl[
-({t_2}{}-{t_3}{}) \left({t_2}''{}-{t_3}''{}\right)+{t_2}'{}^2+{t_3}'{ }^2
\Bigr]
\biggr)
=0,
\label{Q23}
\end{equation}
Note that for ${\gamma}(\phi)=0 $ the term with $R^2$ disappears from the Lagrangian, but the Gauss-Bonnet term remains.

Further, we will show that the dependence of the scalar field $\phi$ on the variable $x^1$ ignored null, which is not included in the metric, does not arise in the models under consideration.

\subsection{
Variant I. ${\phi_1}'\ne 0$, ${\gamma}(\phi)\ne 0 $.
}

Let ${\gamma}(\phi)\ne 0 $ and the field equation $Q_{23}=0$ be satisfied. Then we obtain a functional equation for the metric functions $t_2(x^2)$ and $t_3(x^3)$ of the form:
\begin{equation}
{t_2}''' {t_3}' ({t_2}{}-{t_3}{})^2
+{t_3}''' {t_2}'{} ({t_2}{}-{t_3}{})^2
+6 {t_2}'{}{t_3}'
\Bigl(
-({t_2}{}-{t_3}{}) \left({t_2}''{}-{t_3}''{}\right)+{t_2}'{}^2+{t_3}'{ }^2
\bigr)
=0.
\label{Q23I}
\end{equation}
Having carried out the separation of variables in the equation (\ref{Q23I}), we obtain the following relations for the derivatives of the functions $t_2(x^2)$ and $t_3(x^3)$, which turn the equation (\ref{Q23I}) into an identity:
\begin{equation}
\left({t_2}'\right)^2 = {B} {t_2}{}^4+2 {\alpha} {t_2}{}^3+{\beta} {t_2}{}^2
+2 {\gamma}{t_2}{}
+{\delta}
\label{Q23I2}
,
\end{equation}
\begin{equation}
\left({t_3}'\right)^2 = -{B} {t_3}{}^4-2 {\alpha} {t_3}{}^3-{\beta} {t_3}{}^2
-2 {\gamma}{t_3}{}
-{\delta}
\label{Q23I3}
,
\end{equation}
where ${B}$, $\alpha$, $\beta$, $\gamma$ and $\delta$ are constant parameters.

When the relations (\ref{Q23I2}) and (\ref{Q23I3}) hold, the field equations $Q_{12}=0$ and $Q_{13}=0$ give a condition of the form
\begin{equation}
{B}\gamma'(\phi)=0
\label{Q23IQ12}
,
\end{equation}
Let us further consider the arising cases ${B}=0$ and $\gamma'(\phi)=0$ separately.

\subsubsection{
Variant I. A. ${\phi_1}'\ne 0$, $\gamma'(\phi)\ne 0$.
}

In the considered variant ${B}=0$ and the field equations $Q_{23}=0$, $Q_{12}=Q_{13}=0$ and $Q_{02}=Q_{03}=0$ turn into identities.
The equation $Q_{11}=0$ gives a relation of the following form:
$$
{\phi_1}'{}^2
{\phi}{}
\Bigl(
2 {\alpha} {f_0}{} \left({\gamma}''({\phi}{})-2 \kappa ^2 {\xi}''({\phi}{})\right) +{\sigma}''({\phi}{})+{\omega} \kappa ^2
\bigr)
$$
\begin{equation}
\mbox{}
+{\phi_1}{} {\phi_1}''{} \Bigl(2 {\alpha} {f_0}{} \left({\gamma}'({\phi}{})-2 \kappa ^2 {\xi}'({\phi}{})\right)+{\sigma}'({\phi}{})
\bigr)
=0.
\end{equation}
Assuming that ${f_0}'\ne 0$ and ${\phi_1}'(x^1)\ne 0$, in the equation $Q_{11}=0$ we separate the variables $x^0$ and $x^ 1$ and obtain equations for the functions ${\phi_1}(x^1)$, ${\sigma}({\phi})$, ${\gamma}({\phi})$ and ${\xi} ({\phi})$ of the following form
\begin{equation}
{\phi_1}{\phi_1}''=
{K}{\phi_1}'{}^2
,\qquad
{K}=\mbox{const}
\label{I1Phi1}
,
\end{equation}
\begin{equation}
\phi \left({\gamma}''({\phi}{})-2 \kappa ^2 {\xi}''({\phi}{})\right)
+
{K}
\left({\gamma}'({\phi}{})-2 \kappa ^2 {\xi}'({\phi}{})\right)
=0,
\label{I1GammaXi}
\end{equation}
\begin{equation}
\phi ({\sigma}''({\phi}{})+{\omega} \kappa ^2)
+{K}{\sigma}'({\phi}{})
=0
\quad
\to\quad
{\sigma}'({\phi}{})\ne 0
\label{I1Sigma}
.
\end{equation}

On the other hand, from a scalar equation (\ref{ScalarEquation}), which takes the following form:
\begin{equation}
{\omega}{\phi_1}'{}
\left({\phi_0}{} {f_0}'{}-2 {f_0}{} {\phi_0}'{}\right)
-\frac{{\alpha}^2 {f_0}{}^2 }{2 \kappa ^2}
\bigl(
{\gamma}'({\phi})
+
5\kappa ^2{\xi}'({\phi})
\bigr)
-\frac{{\alpha} {f_0}{}
}{2 \kappa ^2}
{\sigma}'({\phi}{})
+{V}'({\phi}{})
=0,
\end{equation}
we get ${f_0}'(x^0)\ne 0$, ${\phi_1}'(x^1)\ne 0$ and ${\sigma}'({\phi}{})\ne 0 $ ratios of the form:
\begin{equation}
{\alpha}{\sigma}'({\phi}{})=
2 \kappa ^2 {\omega}
{\phi_1}'{}
\,
\frac{{\phi_0}{f_0}'-2 {f_0}{\phi_0}'}{{f_0}}
,
\end{equation}
\begin{equation}
{\gamma}'({\phi})=- 5\kappa ^2{\xi}'({\phi})
,
\end{equation}
\begin{equation}
{V}=\mbox{const}
.
\end{equation}

Note that for $\alpha=0$ in the variant under consideration
the scalar curvature and the Gauss-Bonnet term vanish, and the spacetime becomes conformally flat, but the Riemann curvature tensor generally does not vanish, and the metric generally depends on the wave variable $x^0$.

Assuming that $\alpha\ne 0$, from the scalar equation, the equation $Q_{11}=0$ and their compatibility conditions, we obtain:
\begin{equation}
{\phi_1}(x^1)={p}\exp{({q}x^1)}
,\qquad
{K}=1
,
\end{equation}
\begin{equation}
{\gamma}({\phi})=
\mbox{const}
,\qquad
\xi ({\phi})=
\mbox{const}
,\qquad
{V}=\mbox{const}=\Lambda/(2\kappa^2)
.
\end{equation}
\begin{equation}
\sigma ({\phi})=-\frac{\omega\kappa^2}{4}\,\phi^2+\mbox{const}
,
\end{equation}
where ${q}$, ${p}$ and $\Lambda$ are constant parameters.

Since, as a result of solving the field equations, we obtained that ${\gamma}'({\phi})=0$, the variant considered by us in this section has led to a contradiction.

\subsubsection{
Variant I. B. ${\phi_1}'\ne 0$, $\gamma'(\phi)=0$ and ${\gamma}(\phi)\ne 0 $.
}

Let the relations (\ref{Q23I2}) and (\ref{Q23I3}) now hold, but
$\gamma'(\phi)=0$ and $\gamma\ne 0$.
Then it follows from the equations $Q_{02}=0$ and $Q_{03}=0$ that in the relations (\ref{Q23I2}) and (\ref{Q23I3}) it is necessary to set ${B}=0$. Then, in the equation $Q_{11}=0$, separating the variables $x^0$ and $x^1$ leads to the relations:
\begin{equation}
{\phi_1}{} {\phi_1}''={K}{\phi_1}'{}^2
\label{IBphi1DD}
,
\end{equation}
\begin{equation}
{\phi}{\xi}''({\phi}{})
+{K}{\xi}'({\phi}{})
=0
,
\end{equation}
\begin{equation}
{\phi}{}
\left(
{\sigma}''({\phi}{})
+{\omega}\kappa ^2
\right)
+{K}
{\sigma}'({\phi}{})
=0
\label{IBsigma2}
.
\end{equation}
It follows from the last equation that ${\sigma}'({\phi}{})\ne 0$. Then from the scalar equation, separating the variable $x^0$, we obtain the following relations:
\begin{equation}
{\alpha}
{\sigma}'({\phi}{})
=
2{\omega}{\kappa ^2}
{\phi_1}'
\,
\frac{
{\phi_0}{} {f_0}'{}-2 {f_0}{} {\phi_0}'{}
}{{f_0}}
\label{IBsigma1}
,
\end{equation}
\begin{equation}
{\gamma}({\phi})=
\mbox{const}
,\qquad
\xi ({\phi})=
\mbox{const}
,\qquad
{V}=\mbox{const}=\Lambda/(2\kappa^2)
.
\end{equation}
where $\Lambda$ is a constant parameter that plays the role of a cosmological constant.

From the equations (\ref{IBsigma2}) and (\ref{IBsigma1}) and their compatibility conditions, we obtain:
\begin{equation}
\sigma ({\phi})=-\frac{\omega\kappa^2}{4}\,\phi^2+\mbox{const}
,\qquad
{K}=1
.
\end{equation}
Then from the equations (\ref{IBphi1DD}) and (\ref{IBsigma1}) we get
\begin{equation}
{\phi_1}(x^1)={p}\exp{({q}x^1)}
,
\end{equation}
\begin{equation}
{f_0}(x^0)={c}{\phi_0}^2
\exp{
\left(
-\frac{\alpha x^0}{4q}
\right)
}
,
\end{equation}
where ${q}$, ${p}$ and ${c}$ are constant parameters.

Then the equations $Q_{22}=Q_{33}=0$ lead to the condition:
\begin{equation}
{q}{\phi_0}'=\frac{\alpha}{8}\,{\phi_0}
.
\end{equation}
If ${q}\ne 0$, i.e. ${\phi_1}'\ne 0$, then from the equations $Q_{22}=Q_{33}=0$ we get:
\begin{equation}
{\phi_0}(x^0)=\exp{\left(\frac{\alpha }{8{q}}\,x^0\right)}
,
\end{equation}
which leads to the ratio
\begin{equation}
{f_0}(x^0)=\mbox{const}
,
\end{equation}
which means that the metric does not depend on the wave variable $x^0$, i.e. we have obtained a contradiction in the case under consideration.

\subsection{
Variant II. ${\phi_1}'\ne 0$, ${\gamma}(\phi)= 0 $
}

Let us now consider the case when the field equation $Q_{23}=0$ is satisfied due to the condition ${\gamma}(\phi)= 0 $, then the term with $R^2$ disappears from the Lagrangian, but
the Gauss-Bonnet term with coefficient $\xi(\phi)$ remains.

Then, under the condition ${\gamma}(\phi)= 0$, the scalar equation (\ref{ScalarEquation}) takes the form:
\begin{equation}
{\omega} {\phi_1}'{} 
\left({\phi_0}{} {f_0}'{}-2 {f_0}{} {\phi_0}'{}\right)
+
{f_0}{} {\sigma}'({\phi}{}) 
\frac{F(x^2,x^3)}{2 \kappa ^2}
-
\frac{5}{2}
{f_0}{}^2 {\xi}'({\phi}{}) 
\left(F(x^2,x^3)\right)^2
+{V}'({\phi}{})
=0,
\label{IIScalarEq}
\end{equation}
where
\begin{equation}
F(x^2,x^3)=
\frac{ -({t_2}-{t_3}) \left({{t_2}''}-{{t_3}''}\right)+{\left({t_2}'\right)}^2+{\left({t_3}'\right)}^2}{({t_2}-{t_3})^3}
\label{EqF23}
.
\end{equation}
If the function $F(x^2,x^3)$ does not become a constant, then from (\ref{IIScalarEq}) we obtain the following relations:
\begin{equation}
{V}'({\phi}{})=
-
{\omega} {\phi_1}'{} 
\left({\phi_0}{} {f_0}'{}-2 {f_0}{} {\phi_0}'{}\right)
,\qquad
{\sigma}'({\phi}{}) ={\xi}'({\phi}{})=0
.
\end{equation}
If $F(x^2,x^3)=\mbox{const}={\theta}\ne 0$, then ${f_0}'\ne 0$ gives rise to two additional variants:
\begin{equation}
{\sigma}'({\phi}{}) =
-
\frac{2{\omega}\kappa ^2}{{\theta}}
{\phi_1}'{} 
\frac{
\left(
{\phi_0}{} {f_0}'{}-2 {f_0}{} {\phi_0}'{}
\right)
}{f_0}
,\qquad
{\xi}'({\phi}{}) ={V}'({\phi}{})=0
\label{IIsigma}
,
\end{equation}
\begin{equation}
{\xi}'({\phi}{}) =
\frac{2{\omega}}{5{\theta}^2}
{\phi_1}'{} 
\frac{
\left(
{\phi_0}{} {f_0}'{}-2 {f_0}{} {\phi_0}'{}
\right)
}{{f_0}^2}
,\qquad
{\sigma}'({\phi}{}) ={V}'({\phi}{})=0
\label{IIxi}
.
\end{equation}
Finally, the fourth variant arises if $F(x^2,x^3)=0$, when only one relation remains from the scalar equation:
\begin{equation}
{V}'({\phi}{})=
-
{\omega} {\phi_1}'{} 
\left({\phi_0}{} {f_0}'{}-2 {f_0}{} {\phi_0}'{}\right)
\label{IISeparationForV1}
.
\end{equation}
But the variant $F(x^2,x^3)=0$ leads to a conformally flat spacetime,
when the scalar curvature $R$ and the Gauss-Bonnet term vanish. Note that the Riemann curvature tensor in the general case does not vanish in this version, and the metric
  depends on the wave variable. Thus, there are no solutions for the quadratic theory in this version either.

Relations like (\ref{IISeparationForV1}), also appearing in equations
(\ref{IIsigma}) and (\ref{IIxi}) result due to the condition
\begin{equation}
\partial_0\partial_1 \left(\log{ {V}'({\phi}{}) }\right)=0
\end{equation}
to the following form of the equation for the function ${V}({\phi}{}) $:
\begin{equation}
\phi
\left(
\frac{{V}'''}{{V}'}
-
\left(
\frac{{V}''}{{V}'}
\right)^2
\right)
+
\frac{{V}''}{{V}'}
=0
\label{IISeparationForV2}
.
\end{equation}
The solution to the equation (\ref{IISeparationForV2}) is:
\begin{equation}
{V}({\phi}{})=a \phi^b+c
\label{IISolutionForV}
,
\end{equation}
where $a$, $b$ and $c$ are constants.
The functions $\sigma(\phi)$ and $\xi(\phi)$ also have a similar form for the equations
(\ref{IIsigma}) and (\ref{IIxi}).

We will analyze below the specific variants from this section.

\subsubsection{
Variant II. A. ${\phi_1}'\ne 0$, ${\gamma}(\phi)= 0$, $F(x^2,x^3)\ne \mbox{const}$
 }

Assuming that $F(x^2,x^3)\ne \mbox{const}$, from the scalar equation we obtain the relations:
\begin{equation}
{\gamma}(\phi)= 0
,\qquad
{\sigma}({\phi}{}) =\mbox{const}
,\qquad
{\xi}({\phi}{})=\mbox{const}
,
\end{equation}
\begin{equation}
{V}'({\phi}{})=
-
{\omega} {\phi_1}'{} 
\left({\phi_0}{} {f_0}'{}-2 {f_0}{} {\phi_0}'{}\right)
.
\end{equation}
Thus, the Gauss-Bonnet term in the Lagrangian of the theory has a constant coefficient in this version and does not affect the dynamics of the model. But we will consider this variant for completeness.
Taking into account the solution (\ref{IISolutionForV}), we get:
\begin{equation}
{\phi_1}'
(\phi_1)^{1-b}
\left(-{\phi_0}{} {f_0}'{}+2 {f_0}{} {\phi_0}'{}\right)
(\phi_0)^{1-b}
=
\frac{a b}{{\omega}}
.
\end{equation}
Then we get the following relations:
\begin{equation}
\left(-{\phi_0}{} {f_0}'{}+2 {f_0}{} {\phi_0}'{}\right)
(\phi_0)^{1-b}
=\mbox{const}
={p}
,
\end{equation}
\begin{equation}
{\phi_1}'
(\phi_1)^{1-b}
=\mbox{const}
={q}
,\qquad
{p}{q}=
\frac{a b}{{\omega}}
.
\end{equation}
Then, setting ${a}={p}{\omega}$ and ${b}={q}$, we get:
\begin{equation}
{V}({\phi}{})={\omega}{p} \phi^{q}+\Lambda/(2\kappa^2)
\label{IISolutionForV2}
,
\end{equation}
\begin{equation}
{f_0}'
=-{p}(\phi_0)^{q-2}+2 {f_0}{\phi_0}'/{\phi_0}
,
\end{equation}
\begin{equation}
{\phi_1}'
=
q
(\phi_1)^{q-1}
.
\end{equation}
Substituting the obtained expressions turns the scalar equation into an identity, but the field equation $Q_{11}=0$ gives the condition:
\begin{equation}
{q}=0
\quad\to\quad
\frac{\partial\phi}{\partial x^1}=0
.
\end{equation}
Thus, this variant leads to a contradiction.

\subsubsection{
Variant II. B. ${\phi_1}'\ne 0$, ${\gamma}(\phi)= 0$, $F(x^2,x^3)= \mbox{const}\ne 0$,
${\sigma}'({\phi}) \ne 0$.
}

This variant leads us to an equation of the form
\begin{equation}
F(x^2,x^3)=
\frac{ -({t_2}-{t_3}) \left({{t_2}''}-{{t_3}''}\right)+{\left({t_2}'\right)}^2+ {\left({t_3}'\right)}^2}{({t_2}-{t_3})^3}
=\mbox{const}={\theta}\ne 0
\label{EqF23const}
.
\end{equation}
If $F(x^2,x^3)=\mbox{const}={\theta}\ne 0$, then due to ${f_0}'\ne 0$, the scalar equation implies two possible variants in this case:
\begin{equation}
{\sigma}'({\phi}{}) =
-
\frac{2{\omega}\kappa ^2}{{\theta}}
{\phi_1}'{} 
\frac{
\left(
{\phi_0}{} {f_0}'{}-2 {f_0}{} {\phi_0}'{}
\right)
}{f_0}
,\qquad
{\xi}'({\phi}{}) ={V}'({\phi}{})=0
\label{IIsigma2}
,
\end{equation}
\begin{equation}
{\xi}'({\phi}{}) =
\frac{2{\omega}}{5{\theta}^2}
{\phi_1}'{} 
\frac{
\left(
{\phi_0}{} {f_0}'{}-2 {f_0}{} {\phi_0}'{}
\right)
}{{f_0}^2}
,\qquad
{\sigma}'({\phi}{}) ={V}'({\phi}{})=0
\label{IIxi2}
.
\end{equation}

For both emerging variants, it is necessary to take into account the equation (\ref{EqF23const}), which is a functional equation
on the metric functions $t_2(x^2)$ and $t_3(x^3)$. The solution of the equation, although rather cumbersome, does not cause fundamental difficulties and leads to the following ordinary differential equations for the functions $t_2(x^2)$ and $t_3(x^3)$:
\begin{eqnarray}
\left({t_2}'\right)^2 &=&
2\alpha \left({t_2}\right)^3+\beta \left({t_2}\right)^2+2\gamma t_2+\delta
,\label{Eqt2}\\
\left({t_3}'\right)^2 &=&
-\Bigl(
2\alpha \left({t_3}\right)^3+\beta \left({t_3}\right)^2+2\gamma t_3+\delta
\bigr)
,\label{Eqt3}
\end{eqnarray}
where $\alpha$, $\beta$, $\gamma$ and $\delta$ are constant parameters.

Accordingly, for the second derivatives we have the expressions:
\begin{eqnarray}
{t_2}''
&=&
\alpha \left({t_2}\right)^2+\beta {t_2}+\gamma
,\\
{t_3}''
&=&
-\Bigl(
\alpha \left({t_3}\right)^2+\beta {t_3}+\gamma
\bigr)
.
\end{eqnarray}
Consider in this section the variant (\ref{IIsigma2}), then using the solution (\ref{IISolutionForV}), we obtain
\begin{equation}
{\sigma}({\phi}{})=
one
-
\frac{2{\omega}\kappa ^2}{{\alpha}}
{p} \phi^{q}
\label{IISolutionForSigma2}
,\qquad
{V}({\phi}{})=\mbox{const}=\Lambda/(2\kappa^2)
,\qquad
{\xi}({\phi}{})=\mbox{const}
,
\end{equation}
where
\begin{equation}
{f_0}'
=
{f_0}
\left(
2 {\phi_0}'/{\phi_0}-{p}(\phi_0)^{q-2}
\right)
,
\end{equation}
\begin{equation}
{\phi_1}'
=
q
(\phi_1)^{q-1}
.
\end{equation}
Using the obtained relations, from the field equation $Q_{11}=0$ we obtain conditions of the form:
\begin{equation}
{q}=2
,\qquad
\alpha=8{p}
.
\end{equation}
Then from the field equations $Q_{22}=Q_{33}=0$ we obtain the conditions
\begin{equation}
{V}=\Lambda=0
,\qquad
\xi=0
,
\end{equation}
\begin{equation}
{\phi_0}'=-\frac{{p}}{2}\,{\phi_0}
\quad\to\quad
{\phi_0}=\exp{\left(-\frac{{p}}{2}\,x^0\right)}
.
\end{equation}
Using the obtained relations, from the field equation $Q_{01}=0$ we obtain the condition
\begin{equation}
{p}=0
\quad\to\quad
\frac{\partial\phi}{\partial x^0}=0
,
\end{equation}
which leads us to a contradiction in this variant.

\subsubsection{
Variant II. C. ${\phi_1}'\ne 0$, ${\gamma}(\phi)= 0$, $F(x^2,x^3)= \mbox{const}\ne 0$, 
${\xi}'({\phi}) \ne 0$.
}

Consider the following variant, when the scalar equation leads to conditions of the form:
\begin{equation}
{\xi}'({\phi}{}) =
\frac{2{\omega}}{5{\alpha}^2}
{\phi_1}'{} 
\frac{
\left(
{\phi_0}{} {f_0}'{}-2 {f_0}{} {\phi_0}'{}
\right)
}{{f_0}^2}
,\qquad
{\sigma}
=1
,\qquad
{V}=\mbox{const}=\Lambda/(2\kappa^2)
\label{IIxi3}
.
\end{equation}
Using for ${\xi}({\phi})$ a solution of the type (\ref{IISolutionForV}), we obtain from (\ref{IIxi3}) a relation of the following form:
\begin{equation}
\xi(\phi)=
\frac{2{\omega}}{5{\alpha}^2}
p\phi^q
,
\end{equation}
\begin{equation}
{f_0}'
=
{f_0}
\left(
2 {\phi_0}'/{\phi_0}-{p}{f_0}(\phi_0)^{q-2}
\right)
,
\end{equation}
\begin{equation}
{\phi_1}'
=
q
(\phi_1)^{q-1}
.
\end{equation}
Using the obtained relations, from the field equation $Q_{11}=0$ we obtain
\begin{equation}
{q}=1
,\qquad
\alpha=0
.
\end{equation}

The condition $\alpha=0$ leads to a contradiction, because the function $F_{23}$ vanishes in this case. The variant leads to a conformally flat spacetime with zero scalar curvature $R$ and zero Gauss-Bonnet term. The Riemann curvature tensor generally does not vanish, and the metric
 depends on the wave variable. Thus, there are no solutions for the quadratic theory in this version.
 
Thus, in the analysis of variant I and variant II, we have shown that the dependence of the scalar field on the  null variable $x^1$ leads to contradictions and the absence of solutions of this type.

\subsection{Variant III. ${\phi}={\phi}(x^0)$.}

As shown above, the scalar field does not depend on the null variable $x^1$.
Let us now consider the case when the scalar field depends on the wave variable $x^0$.
The equation $Q_{11}=0$ gives the condition $L_1=0$, which in turn, using the normalization condition, gives $L_2=L_3=0$.

From the equation $Q_{23}=0$ we obtain a condition similar to variant I:
$$
{\gamma}({\phi}{}) 
\biggl(
6 {t_2}'{t_3}' \left(-({t_2}{}-{t_3}{}) \left({t_2}''{}-{t_3}''{}\right)+{t_2}'{}^2+{t_3}'{}^2\right)
$$
\begin{equation}
\mbox{}
+{t_2}^{(3)}{t_3}' ({t_2}{}-{t_3}{})^2
+{t_3}^{(3)}{} {t_2}'{} ({t_2}{}-{t_3}{})^2
\biggr)
=0,
\end{equation}

Let us consider separately two cases when ${\gamma}({\phi}{}) \ne 0$ and when ${\gamma}=0 $.

\subsubsection{Variant III. A. ${\phi}={\phi}(x^0)$, ${\gamma}({\phi}{}) =0$.}

The condition ${\gamma}({\phi}{})=0$ turns the field equation $Q_{23}=0$ into an identity, but means the exclusion of the term with $R^2$ from the Lagrangian. The Gauss-Bonnet term in the Lagrangian is preserved.

The field equations $Q_{22}=Q_{33}=0$ give the following relation:
\begin{equation}
5 {f_0}{}^2 {\xi}({\phi}{})
\left[
\frac{
-({t_2}{}-{t_3}{}) \left({t_2}''{}-{t_3}''{}\right)+{t_2}'{}^2+{t_3}'{ }^2
}{({t_2}{}-{t_3}{})^3}
\right]^2
=
2 {V}({\phi}{})
,
\end{equation}
which allows you to separate variables. As a result, we get that either ${\xi}({\phi}{})={V}({\phi}{}) =0$, or:
\begin{equation}
\frac{
-({t_2}{}-{t_3}{}) \left({t_2}''{}-{t_3}''{}\right)+{t_2}'{}^2+{t_3}'{}^2
}{({t_2}{}-{t_3}{})^3}={\alpha}
,\qquad
{\alpha}=\mbox{const}
\label{IIIeqt2t3}
,
\end{equation}
\begin{equation}
{V}({\phi}{}) =
\frac{
5{\alpha}^2 
}{2}
\,
{f_0}{}^2 
{\xi}({\phi}{}) 
\label{IIIeqV}
,
\end{equation}
A functional equation of the form (\ref{IIIeqt2t3}) has already appeared in variant II. The solution looks like:
\begin{eqnarray}
\left({t_2}'\right)^2 &=&
2\alpha \left({t_2}\right)^3+\beta \left({t_2}\right)^2+2\gamma t_2+\delta
,\label{IIIEqt2}\\
\left({t_3}'\right)^2 &=&
-\Bigl(
2\alpha \left({t_3}\right)^3+\beta \left({t_3}\right)^2+2\gamma t_3+\delta
\Bigr)
,\label{IIIEqt3}
\end{eqnarray}
where $\alpha$, $\beta$, $\gamma$ and $\delta$ are constant parameters.

Substituting the relations (\ref{IIIeqV}), (\ref{IIIEqt2}), and (\ref{IIIEqt3}) into the field equations, we obtain the condition $\alpha=0$ from the equation $Q_{01}=0$. In this case, the only remaining equation $Q_{00}=\varepsilon {L_0}^2$, which did not turn into an identity, gives a relation between the radiation energy density $\varepsilon(x^0)$ and the functions $f_0(x^0)$ , $\phi(x^0)$ and $\sigma(\phi)$:
\begin{equation}
{\kappa ^2}{L_0}^2\varepsilon 
=
\frac{
{f_0}''{} {\sigma}({\phi}{})
-{f_0}'{} {\phi}'{} {\sigma}'({\phi}{})
}{{f_0}{}}
-\frac{{f_0}'{}^2 {\sigma}({\phi}{})}{2{f_0}{}^2}
-{\phi}''{} {\sigma}'({\phi}{})
-{\phi}'{}^2
\left(
 {\sigma}''({\phi}{})
 +{\kappa ^2}{\omega}
 \right)
\label{IIIenergy1}
,
\end{equation}
The resulting exact solution is conformally flat with a nonzero Riemann curvature tensor, and the metric and two independent components of the Riemann curvature tensor depend on the wave variable $x^0$. Note that the exact solution obtained here leads to zeroing of the quadratic terms in the action.

Assuming additionally $\sigma=1$, we obtain the exact solution for a plane gravitational wave in Einstein's theory of gravity with pure radiation having energy density $\varepsilon(x^0)$ and wave vector $L_\alpha=\{L_0,0 ,0,0\}$ and a scalar field $\phi(x^0)$, which are related by a relation of the form:
\begin{equation}
\frac{{f_0}''}{{f_0}{}}
-
\frac{{f_0}'{}^2 }{2{f_0}{}^2}
=
{\kappa ^2}
\left(
{L_0}^2\varepsilon +
{\omega}{\phi}'{}^2
\right)
\label{IIIenergy2}
.
\end{equation}

Note that the vanishing of the right side of the equality (\ref{IIIenergy2}) leads to the degeneration of the space into a flat Minkowski spacetime.
The functions $f_0(x^0)$, $t_2(x^2)$, and $t_3(x^3)$ in the metric correspond to the relationship (\ref{IIIenergy1}) (or (\ref{IIIenergy2}) ) and equations (\ref{IIIEqt2})--(\ref{IIIEqt3}).

Thus, in this section, we have obtained an exact solution for a gravitational wave in a type I Shapovalov space in a theory with a scalar field, but without quadratic terms in the Lagrangian.

\subsubsection{Variant III. B. ${\phi}={\phi}(x^0)$, ${\gamma}({\phi}{}) \ne 0$.}

For ${\gamma}({\phi}{}) \ne 0$, from the equation $Q_{23}=0$ we obtain a functional equation for the metric functions $t_2(x^2)$ and $t_3(x^3) $, which already appeared in variant I, of the following form:
\begin{equation}
{t_2}^{(3)}{t_3}' ({t_2}{}-{t_3}{})^2
+{t_3}^{(3)}{} {t_2}'{} ({t_2}{}-{t_3}{})^2
+
6 {t_2}'{t_3}'
\Bigl(
-({t_2}{}-{t_3}{}) \left({t_2}''{}-{t_3}''{}\right)+{t_2}'{}^2+{t_3}'{ }^2
\bigr)
=0,
\label{IIIBt2t3}
\end{equation}
Substitution of solutions (\ref{Q23I2})--(\ref{Q23I3}) of equation (\ref{IIIBt2t3}) into the remaining field equations leads to solutions of the form (\ref{IIIEqt2})--(\ref{IIIEqt3}) . As a result, the field equations $Q_{22}=Q_{33}=0$ give a condition of the form:
\begin{equation}
{V}({\phi})=
\frac{
{\alpha}^2 {f_0}^2
}{2 \kappa ^2}
\left(
5 \kappa^2 {\xi}({\phi})-{\gamma}({\phi})
\right)
\label{IIIBV}
.
\end{equation}
Then from the equation $Q_{01}=0$ we get:
\begin{equation}
{\alpha} 
\Bigl(
{\sigma}({\phi}{})
+
2{\alpha} {f_0}{}
   {\gamma}({\phi}{}) 
\Bigr)
=0
\label{IIIBsigma}
.
\end{equation}

Assuming that ${\alpha}\ne 0$ and using the relations (\ref{IIIBV})--(\ref{IIIBsigma}), we obtain the following condition from the scalar equation:
\begin{equation}
{\alpha}\,{f_0}'(x^0)\,{\xi}({\phi})=0
.
\end{equation}
For $\alpha=0$ we get the conformally flat solution considered earlier, for ${f_0}'(x^0)$ the metric ceases to depend on the wave variable, so we consider the case of ${\xi}({\phi} )=0$.
Assuming that the coefficient at the Gauss-Bonnet term vanishes (${\xi}=0$), from the only remaining field equation $Q_{00}=\varepsilon {L_0}^2$ we obtain a relation of the form:
\begin{equation}
{{\phi}'}^2 (x^0)
=
-
\frac{{L_0}^2
}{{\omega}}
\,
{\varepsilon}(x^0)
\label{IIIBenergyQ00}
.
\end{equation}
The relation (\ref{IIIBenergyQ00}) relates the pure radiation energy density function ${\varepsilon}(x^0)$ (the wave vector has the form $L_\alpha=\{L_0,0,0,0\}$) with the function scalar field $\phi(x^0)$.
The function $\gamma(\phi)$ with the quadratic term $R^2$ in the Lagrangian is related to the function $\sigma(\phi)$ with the term $R$ and with the conformal factor of the metric ${f_0}(x^0)$ relation (\ref{IIIBsigma}), which takes the following form:
\begin{equation}
\sigma(\phi)=-2\alpha f_0 \gamma(\phi)
.
\end{equation}

The scalar potential function $V(\phi)$ is defined by the relation:
\begin{equation}
{V}({\phi})=
-
\frac{\alpha^2}{2\kappa^2}
\,
{f_0}^2
\,
\gamma(\phi)
\label{IIIBV2}
.
\end{equation}
The metric functions $t_2(x^2)$ and $t_3(x^3)$ are defined by the equations (\ref{IIIEqt2})--(\ref{IIIEqt3}).

For $\alpha=0$ the obtained solution passes into a conformally flat space with zero scalar curvature and zero Gauss-Bonnet term, and the Riemann curvature tensor generally does not vanish and depends, like the metric, on the wave variable $x^ 0$.

Thus, we have obtained a solution for a plane gravitational wave in the quadratic theory of gravity with a scalar field and
radiation.

\section{
The eikonal equation and trajectories of test particles
}

The eikonal equation ($\Psi$ is the eikonal function)  determines the propagation of the wave front in
spacetime in the high-frequency approximation and has the form:
\begin{equation}
g^{\alpha\beta}\frac{\partial \Psi}{\partial x^\alpha}\frac{\partial \Psi}{\partial x^\beta}=0
\label{EikonalEq}
.
\end{equation}

In the type I Shapovalov wave spacetime under consideration, we can construct the eikonal function $\Psi$ by separation of variables, assuming that
\begin{equation}
S={\vartheta}_0(x^0)+{\vartheta}_1(x^1)+{\vartheta}_2(x^2)+{\vartheta}_3(x^3),
\end{equation}
where for the null variable $x^1$, which is not included in the metric, ${\vartheta}_1(x^1)$ can be reduced to the form
${\vartheta}_1={\lambda_1} x^1$,  ${\lambda_1}=\mbox{const}$.

\begin{equation}
\Psi=\psi_0(x^0)+\psi_1(x^1)+\psi_2(x^2)+\psi_3(x^3)
,
\end{equation}
then the equation (\ref{EikonalEq}) gives a relation of the following form:
\begin{equation}
2{\psi_0}'{\psi_1}'( t_3(x^3)-t_2(x^2))+\left({\psi_2}'\right)^2+\left({\psi_3}'\right)^2=0
.
\end{equation}
By an admissible transformation of variables, we can set $\psi_0(x^0)=px^0$ and $\psi_1(x^1)=qx^1$, and then, separating the variables, we obtain
\begin{equation}
\psi_2(x^2)=\int{\sqrt{2pq\, t_2(x^2)-r}}\,dx^2
,\qquad
\psi_3(x^3)=\int{\sqrt{-2pq\, t_3(x^3)+r}}\,dx^3
,\qquad
p,q,r=\mbox{const}
,
\end{equation}
where the metric functions $t_2(x^2)$ and $t_3(x^3)$ are determined by solving the field equations in terms of the relations (\ref{IIIEqt2}) and (\ref{IIIEqt3}).

Similarly, one can determine the trajectories of test particles in the spacetimes under consideration by integrating the Hamilton-Jacobi equation (\ref{HJE}) by the method of separation of variables.

The action function for the test particle in the Hamilton-Jacobi equation (\ref{HJE}) can be written in the ''separated'' form:

Separating the variables in the Hamilton-Jacobi equation (\ref{HJE}) and setting $c=1$ here, we get:
\begin{equation}
\frac{m^2}{f_0(x^0)}-2{\lambda_1}{{\vartheta}_0}'(x^0)=\frac{1}{t_3(x^3)-t_2(x^2)}
\Bigl(
\left( {{\vartheta}_2}'(x^2) \right)^2
+
\left( {{\vartheta}_3}'(x^3) \right)^2
\Bigr)
=\mbox{const}={\lambda_2}
.
\end{equation}
and, further, we obtain relations of the form:
\begin{equation}
\left( {{\vartheta}_2}'(x^2) \right)^2
+{\lambda_2} t_2(x^2)
=
-\left( {{\vartheta}_3}'(x^3) \right)^2
+{\lambda_2} t_3(x^3)
=\mbox{const}={\lambda_3}
.
\end{equation}
Here ${\lambda_1}$, ${\lambda_2}$ and ${\lambda_3}$ are constant parameters determined by the initial or boundary conditions.

Thus, we have obtained the following expression for the test particle action function:
$$
S={\lambda_1}x^1-\frac{{\lambda_2}}{2{\lambda_1}}x^0+\frac{m^2}{2{\lambda_1}}\int{\frac{dx^0}{f_0(x^0)}}
+\varepsilon_2\int{\sqrt{{\lambda_3}-{\lambda_2}t_2(x^2)}\,dx^2}
$$
\begin{equation}
\mbox{}
+\varepsilon_3\int{\sqrt{{\lambda_2}t_3(x^3)-{\lambda_3}}\,dx^2}
,\qquad
\varepsilon_2,\varepsilon_3=\pm 1
.
\end{equation}
Here the functions $t_2(x^2)$ and $t_3(x^3)$ are determined by solving the field equations through the relations (\ref{IIIEqt2}) and (\ref{IIIEqt3}).

The form of particle trajectories will be determined by the relations:
\begin{eqnarray}
\frac{\partial S}{\partial {\lambda_1}}
={\rho}_1
&\to&
x^1+\frac{{\lambda_2}}{2{\lambda_1}^2}x^0+\frac{m^2}{2{\lambda_1}^2}\int{\frac{dx^0}{f_0(x^0)}}={\rho}_1
,\qquad
{\lambda_1}\ne 0
\label{Trajectory1}
,\\
\frac{\partial S}{\partial {\lambda_2}}
={\rho}_2
&\to&
-\frac{x^0}{2{\lambda_1}}
-\frac{\varepsilon_2}{2}\int{\frac{t_2(x^2)\,dx^2}{\sqrt{{\lambda_3}-{\lambda_2}t_2(x^2)}}}
+\frac{\varepsilon_3}{2}\int{\frac{t_3(x^3)\,dx^3}{\sqrt{{\lambda_2}t_3(x^3)-{\lambda_3}}}}
={\rho}_2
\label{Trajectory2}
,\\
\frac{\partial S}{\partial {\lambda_3}}
={\rho}_3
&\to&
\frac{\varepsilon_2}{2}\int{\frac{dx^2}{\sqrt{{\lambda_3}-{\lambda_2}t_2(x^2)}} }
-
\frac{\varepsilon_3}{2}\int{\frac{dx^3}{\sqrt{{\lambda_2}t_3(x^3)-{\lambda_3}}}}
={\rho}_3
.
\label{Trajectory3}
\end{eqnarray}
Here the quantities ${\rho}_1$, ${\rho}_2$ and ${\rho}_3$ are additional constants determined by the initial conditions.

The proper time of the test particle $\tau$ can be represented in the following form:
\begin{equation}
\tau=S\,\bigr|_{m=1}=2{\lambda_1}x^1+\frac{{\lambda_2}}{{\lambda_1}}\,x^0
.
\label{taux0x1}
\end{equation}
From the equation (\ref{Trajectory1}), using the relation (\ref{taux0x1}), we obtain the relationship between the wave variable $x^0$ and the proper time $\tau$ on the trajectories of test particles:
\begin{equation}
\tau=
-
\frac{1}{{\lambda_1}}\int{\frac{dx^0}{f_0(x^0)}}+2{\lambda_1}{\rho}_1
\label{taux0}
.
\end{equation}
The relations obtained completely determine the trajectories of motion of test particles in the considered models of gravitational waves.

\section{Conclusion}

Exact solutions of the equations of the quadratic theory of gravity with a scalar field for Shapovalov type I wave spacetimes are found. Gravitational-wave solutions are obtained, depending on the maximum possible number of variables for wave metrics of 4-dimensional spacetime
(three variables, including the wave variable) in privileged coordinate systems, where it is possible to separate the wave variables in the Hamilton-Jacobi equation for test particles and in the eikonal equation for radiation. The situation is shown to differ from the general theory of relativity, where these gravitational-wave models based on Einstein's vacuum equations degenerate. 
Solutions for the eikonal equation and a general form of test particle trajectories are found for the considered wave models of spacetime. Thus, Shapovalov's wave spacetimes provide an additional mathematical tool for obtaining exact models of gravitational waves, which makes it possible to study possible differences in modified gravity theories and form tests for observational verification of these differences.


%

\authorcontributions{Conceptualization, K.O.; methodology, K.O. and I.K.; validation, I.K., A.F. and K.O.;  investigation, K.O., I.K. and A.F.; writing---original draft preparation, K.O.; supervision, K.O.; project administration, K.O.; funding acquisition, K.O. All authors have read and agreed to the published version of the manuscript.}

\funding{The reported study was funded by RFBR, project number N~20-01-00389~A.}

\conflictsofinterest{The authors declare no conflict of interest. }

\institutionalreview{Not applicable.}

\informedconsent{Not applicable.}

\dataavailability{Not applicable.} 

%

\reftitle{References}

\end{document}